\documentclass[
 reprint,
 amsmath,amssymb,
 aps,prb
]{revtex4-1}

\usepackage{graphicx}
\usepackage{dcolumn}
\usepackage{bm}
\usepackage[normalem]{ulem}
\usepackage{amsmath}

\usepackage{subfigure}

\usepackage{soul,xcolor}
\setstcolor{red}

\RequirePackage{color}
\definecolor{MyDarkGreen}{rgb}{0.02,0.60,0.06}
\definecolor{burgundy}{rgb}{0.5, 0.0, 0.13}

\newcommand{\stkout}[1]{\ifmmode\text{\sout{\ensuremath{#1}}}\else\sout{#1}\fi}
\makeatother
\def\RealP#1{\operatorname{Re} \left({#1}\right)}

\begin{document}

\title{The wiggly cosmic string as a waveguide for massless and massive fields}

\author{Frankbelson dos S. Azevedo}
\email{dossanto2@univ-lorraine.fr}
\affiliation{Statistical Physics Group and Coll\`ege Doctoral ${\mathbb{L}}^4$ for Physics of Complex Systems, IJL, UMR Universit\'e de Lorraine - CNRS 7198, 
 54506 Vand\oe uvre les Nancy, France}

\author{Fernando Moraes}
\email{fernando.jsmoraes@ufrpe.br}
\affiliation{Departamento de F\'{\i}sica, Universidade Federal Rural de Pernambuco, 
52171-900, Recife, PE, Brazil}

\author{Francisco Mireles}
\email{fmireles@cnyn.unam.mx}
\affiliation{Centro de Nanociencias y Nanotecnolog\'\i a, Universidad Nacional Autonoma de M\'exico, Apdo. Postal 14, 
22800 Ensenada B.C., Mexico}

\author{Bertrand Berche}
\email{bertrand.berche@univ-lorraine.fr}
\affiliation{Statistical Physics Group, Laboratoire de Physique et Chimie Th\'eoriques, Universit\'e de Lorraine, 
 54506 Vand\oe uvre les Nancy, France}
\author{S\'ebastien Fumeron}
\email{sebastien.fumeron@univ-lorraine.fr}
\affiliation{Statistical Physics Group, Laboratoire de Physique et Chimie Th\'eoriques, Universit\'e de Lorraine, 
 54506 Vand\oe uvre les Nancy, France}

\date{\today}

\begin{abstract}
We examine the effect of a wiggly cosmic string {for} both  massless and massive particle propagation along the string axis. {We show that the  wave equation that governs the propagation of a scalar field in the neighborhood of a wiggly string is formally equivalent to the quantum wave equation describing the hydrogen atom in two dimensions.  We further show that the wiggly string spacetime behaves as a gravitational waveguide in which the {quantized} wave modes propagate with frequencies that depend on the mass, string energy density,  and string tension. We propose an analogy with an optical fiber, defining an effective refractive index likely to mimic the cosmic string effect in the laboratory.}  
\end{abstract}

\maketitle

\section{Introduction}

The thermal history of the Universe started 14 billion years ago with an extremely hot and dense quark-gluon plasma that cooled down in the inflation era. As a consequence, it has undergone a succession of phase transitions involving spontaneous symmetry-breaking (SSB) {mechanisms}. Below an energy scale $M_{\rm GUT}\sim 10^{16}$ GeV, the strong forces are represented by the three-fold color symmetry, associated with the gauge group $SU(3)_{\rm color}$, whereas {the} weak {and electromagnetic forces} are mixed into the electroweak interaction, represented by weak isospin symmetry (gauge group $SU(2)_{\rm L}$) and hypercharge symmetry (gauge group $U(1)_{\rm Y}$) {\cite{weinberg}}.  This is  the realm of the particle physics standard model, which has been tested to a very high precision. Above $M_{\rm GUT}$, strong and electroweak interactions unify within a larger gauge symmetry group \textit{G} {where} grand unified theories involving supersymmetry (SUSY GUT) {have been} considered as suitable description for such energy scales {\cite{bajc04,fukuyama05,raby11}}. 

As well-known in condensed matter physics (CMP), spontaneoulsy symmetry breaking (SSB) lead to phase transitions which often give rise to the appearance of topological defects. Interestingly, a mechanism originally proposed by Kibble \cite{Kibble1976} to describe the birth and dynamics of a network of defects in a cosmological context revealed to be relevant in the condensed matter realm, for example in liquid crystals. \cite{bowick1994cosmological}.
To determine what kind of topological defect {emerges} for a given SSB transition $G\rightarrow H$, one may study the content of homotopy groups $\pi_k(G/H)$ of the vacuum manifold $\mathcal{M}=G/H$ \cite{Kenna}. If $\pi_k(G/H)\neq 0$, defects of dimension $2-k$ are formed: for $k=0$, defects are 2D (grain boundaries in CMP, domain walls in cosmology), for $k=1$, defects are line-like (disclinations or dislocations in CMP, cosmic strings in cosmology) and for $k=2$, defects are point-like (e.g. hedgehogs in CMP, monopoles in cosmology...). 

Even though the SUSY extension to the standard model  still has to pass experimental verification (the first run of the LHC found no evidence for supersymmetry), it provides a route to the formation of cosmic strings. In a seminal paper, Jeannerot et al \cite{Jeannerot2003} examined all possible SSB patterns from the large possible SUSY GUT gauge groups down to the standard model $SU(3)_{\rm color}\times SU(2)_{\rm L} \times U(1)_{\rm Y}$ and concluded that cosmic string formation was unavoidable. {Another possibility for cosmic string generation is brane inflation \cite{dvali1999brane}. Cosmic strings, seen as lower-dimensional D-branes which are one-dimensional in the noncompact directions, may have been abundantly produced by brane collision towards the end of the brane inflationary period \cite{sarangi2002cosmic}.} 

{  Despite all  theoretical justifications for the existence of cosmic strings, the observational evidence  is still feeble and mostly indirect. Nevertheless, the search for cosmic strings is very active and it happens in such diverse fronts as the cosmic microwave background \cite{1475-7516-2017-06-004} and gravitational wave bursts \cite{stott2017gravitational}.   As warned by Copeland and Kibble \cite{copeland2010cosmic}, ``\textit{Both cosmic strings and superstrings are still purely hypothetical objects.
There is no direct empirical evidence for their existence, though there have been
some intriguing observations that were initially thought to provide such evidence,
but are now generally believed to have been false alarms. Nevertheless, there are
good theoretical reasons for believing that these exotic objects do exist, and
reasonable prospects of detecting their existence within the next few years.}" 
}

In this paper, we study the dynamics of particles in the vicinity of a wiggly cosmic string. Regular straight strings are linear defects for which the geometry is globally {that} of a cone and therefore, spacetime is locally flat, except on string axis. {Indeed, for such objects, the line tension $T_0$ exactly matches the energy density per unit length $\mu_0$, such that straight strings do not gravitate. Recent data on the Cosmic Microwave Background collected from PLANCK satellite have not confirmed the existence of these objects yet, but they have set upper boundaries on their mass-energy density {\cite{planck13} $G\mu_0 < 10^{-7}$} ($c=1$). Refined models for cosmic strings may involve small-scale perturbations such as kinks and wiggles \cite{Vilenkin1994}. The presence of wiggles generates a far gravitational field contribution which  may be responsible for an elliptical distortion of the shape of background galaxies {\cite{Dyda2007,Feng2012}} or {for} the accretion of dark energy around the defect\cite{Madrid2006}. Averaging the effect of these perturbations along a string increases the linear mass density $\tilde{\mu}$ and decreases the string tension $\tilde{T}$, {respecting the equation of state \cite{carter1990,Vilenkin1990}$ \tilde{\mu}\:\tilde{T}=\mu_0^2$}, leading the wiggly string to exert a gravitational pulling on neighboring objects. 
In the weak-field approximation, the linearized line element representing the spacetime of a wiggly string oriented along the $z$-axis is given by}\cite{Vachaspati1991,Vilenkin1994}:
\begin{eqnarray}
	ds^2&=&-\left(1+8\varepsilon\ln\left({r}/{r_0}\right)\right)dt^2 
	+dr^2\nonumber+\alpha^2 r^2 d\theta^2 \\&+&\left(1-8\varepsilon\ln\left({r}/{r_0}\right)\right)dz^2. \label{Vilenkinmetric}
\end{eqnarray}
Here, {$\alpha^2=1-4G(\tilde{\mu}+\tilde{T})$, where $4G(\tilde{\mu}+\tilde{T}) \ll 1$  meaning that conical deficit angle $4G\pi(\tilde{\mu}+\tilde{T})$ associated to the string is very small. The parameter $\varepsilon$ is defined as the excess of mass-energy density, $2\varepsilon=G(\tilde{\mu}-\tilde{T})$. It must be emphasized that $G(\tilde{\mu}+\tilde{T})$ and $\varepsilon$ are two independent parameters: the former accounts for the discrepancy between flat and conical geometries, whereas the latter accounts for the discrepancy between straight and wiggly strings. The constant $r_0$ denotes the effective string radius \cite{patrick1994}.} In the remainder of this work, propagation of particles will be considered only within the region $r_0 < r \ll r_0 e^{1/{8\varepsilon}}$, in order to avoid the logarithmic divergence at small and large distances from the defect. Hence, $h_{00}=8\varepsilon\ln\left({r}/{r_0}\right)=O(\varepsilon) \ll 1$.

In the next sections of this work, the wave equation for propagation along the string axis is numerically solved in the background spacetime given by metric (\ref{Vilenkinmetric}). {The} properties of the radially bound states   and the dispersion relations are examined in detail {for both massless and massive particles}. Then an analogy with light propagation in an optical fiber is performed to design a system likely to mimic the effect of a cosmic wiggly string in laboratory. 

\section{Massless particle propagation}

In a plane perpendicular to the wiggly string, light propagates in the same way as in the vicinity of a regular straight string \cite{Vilenkin1994}, that is without experiencing any gravitational force. On the contrary, when the direction of propagation is not perpendicular to the string, the wiggly string exerts gravitational pulling on light passing by, as the background spacetime is not locally Euclidean.  Consequently, the wave equation governing propagation of a scalar field in this background geometry needs to account for its curvature. This is done by using the 4-dimensional Laplace-Beltrami operator in the wave equation
\begin{eqnarray}
\frac{1}{\sqrt{-g}}\partial_\mu(\sqrt{-g}g^{\mu\nu}\partial_\nu)\Phi=0, \label{waveequation}
\end{eqnarray}
where $\Phi=\Phi(r,\theta,z,t)$ is the scalar wave amplitude  and $g=\det (g_{\mu\nu})$ with the metric tensor $g_{\mu\nu}$ coming from metric (\ref{Vilenkinmetric}). {In terms of the metric (\ref{Vilenkinmetric}), this gives}
\begin{eqnarray}
&&-(1-h_{00})\partial_t^2 \Phi+\frac{1}{r} \partial_r\left(r\partial_r\right) \Phi+\frac{1}{\alpha^2r^2} \partial_{\theta}^2 \Phi \nonumber \\
&&\qquad\qquad+(1+h_{00}) \partial_z^2 \Phi=0. \label{waveequation0} \label{intermediate}
\end{eqnarray}
{As the field is single-valued, $\Phi$ has to be periodic in $\theta$:}
\begin{eqnarray}
\Phi(\theta)=\Phi(\theta+2 \pi).
\label{p}
\end{eqnarray}
To solve equation (\ref{waveequation0}) we make the ansatz 
\begin{eqnarray}
\Phi(r,\theta,z,t)=  e^{i  l \theta} e^{i(\omega t-kz)} R(r), \label{putsolution0}
\end{eqnarray}
where the wave vector $k \in \mathbb{R}$, $l=0,\pm1,\pm2...$ specifies the angular momentum and $\omega$ is an angular frequency. Substituting the general solution (\ref{putsolution0}) into equation (\ref{waveequation0}), 
we get
\begin{eqnarray}
-\frac{1}{r} \frac{d}{d r}\left(r\frac{d R}{d r}\right)&&+\frac{l^2}{\alpha^2 r^2} R+h_{00}\left(\omega^2+k^2\right)R \nonumber \\ 
&&=\left(\omega^2-k^2\right)R.  \label{intermediate}
\end{eqnarray}
Defining the  dimensionless variables $\rho=r/\gamma$, $\rho_0=r_0/\gamma$,  where  
$\gamma=\left[8\varepsilon\left(\omega^2+k^2\right)\right]^{-1/2}$, then multiplying (\ref{intermediate}) by $\gamma^2\sim O(\varepsilon^{-1})$ and rearranging terms gives the eigenvalue equation: 
\begin{eqnarray}
-\frac{1}{\rho}\frac{d}{d\rho}\left(\rho\frac{dR}{d\rho}\right)&+&\left(\frac{{l}^2}{\alpha^2\rho^2}+\ln{\frac{\rho}{\rho_0}}\right)R =\bar{\zeta}R \label{change}
\end{eqnarray}
with
\begin{eqnarray}
\bar{\zeta}=\frac{1}{8\varepsilon}
\frac{\omega^2 -k^2}{\omega^2+k^2}.\label{relation1}
\end{eqnarray}

\begin{figure}[h]
\includegraphics[scale=0.6]{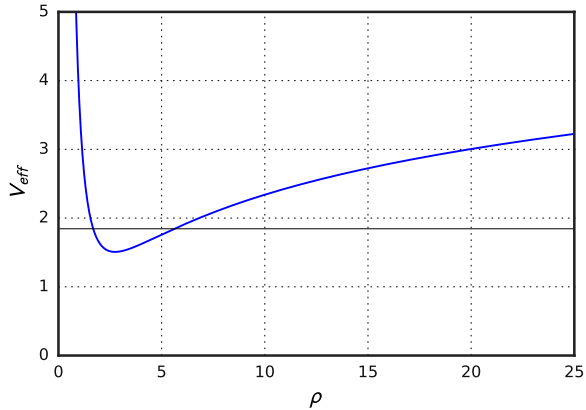}
  \caption{Effective potential, {$V_{\rm eff}$} (in units where $\rho_0=1$), for $l=2$. The horizontal solid line represents the ground state $\bar\zeta$ at that value of $l$.   }
      \label{pot}
\end{figure}

\begin{figure}[h]

\center
\subfigure[ref1][]{\includegraphics[scale=0.5]{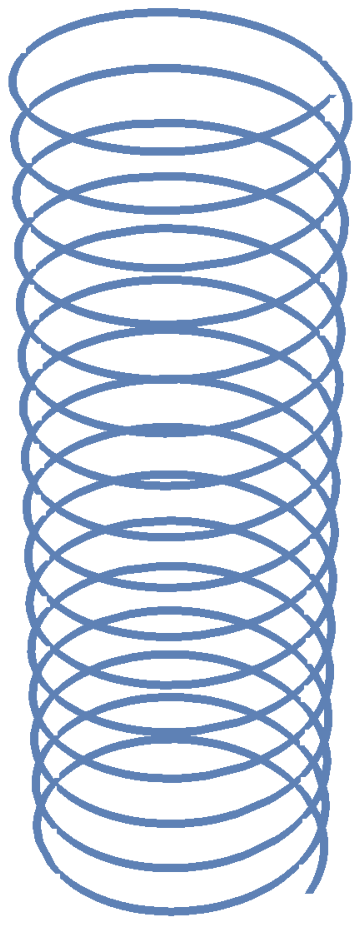}}
\qquad
\subfigure[ref2][]{\includegraphics[scale=0.5]{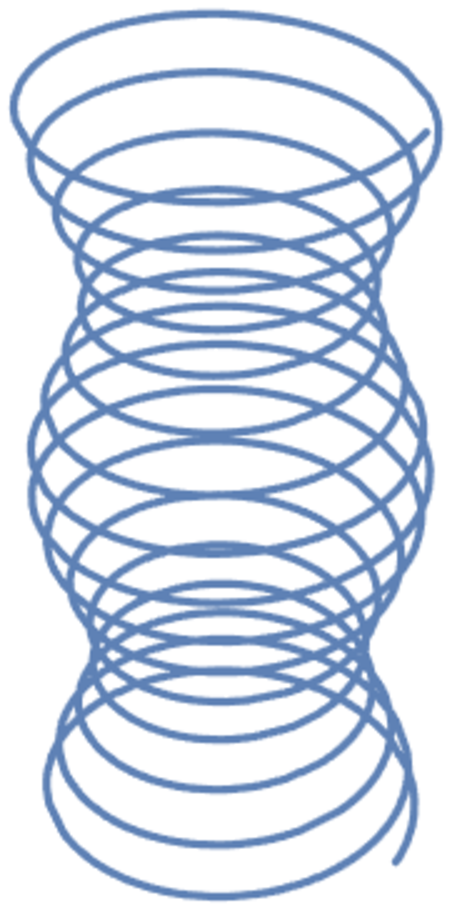}}
\caption{Possible ray paths corresponding to the geometric optics limit of a scalar wave propagating along the wiggly string: (a) when the ``total energy" $\bar{\zeta}$ is at the minimum of the effective potential and (b) at some point above it.}
       \label{trajectories}

\end{figure}

We note that the potential behaving logarithmically, the energy scale cannot be fixed at infinity and we work in the following with $\omega-$dependent length units such that $\rho_0=1$. {Eq. (\ref{change}) is formally equivalent to the Schr\"odinger equation that describes the hydrogen atom in the 2D Coulomb potential. Hence, the potential term in Eq. (\ref{change}), {$V_{\rm eff}=\frac{{l}^2}{\alpha^2 \rho^2}+\ln{\frac{\rho}{\rho_0}}$}}  (see figure \ref{pot}) only accommodates for bound states \cite{Austria1985,Eveker1990,Garon2013}.
As a consequence, in the geometrical {optics} limit, trajectories are radially bounded helices around the string, as appears in Fig. \ref{trajectories}, {explicitly showing the gravitational pulling by the string}. This is in agreement with {Ref. \onlinecite{Arazi2000}}, where geodesics near a Brans-Dicke wiggly cosmic string were also found to be bounded. The minimum and maximum radii are solutions of the transcendental equation $\bar{\zeta}=V_{\rm eff}$ whereas the pitch is given by the ratio between the angular and effective linear momenta ${\frac{l}{k}}$. In the case of $l=0$  the trajectory is rectilinear and parallel to the string. 

\begin{figure}[h] 
	\includegraphics[scale=0.6]{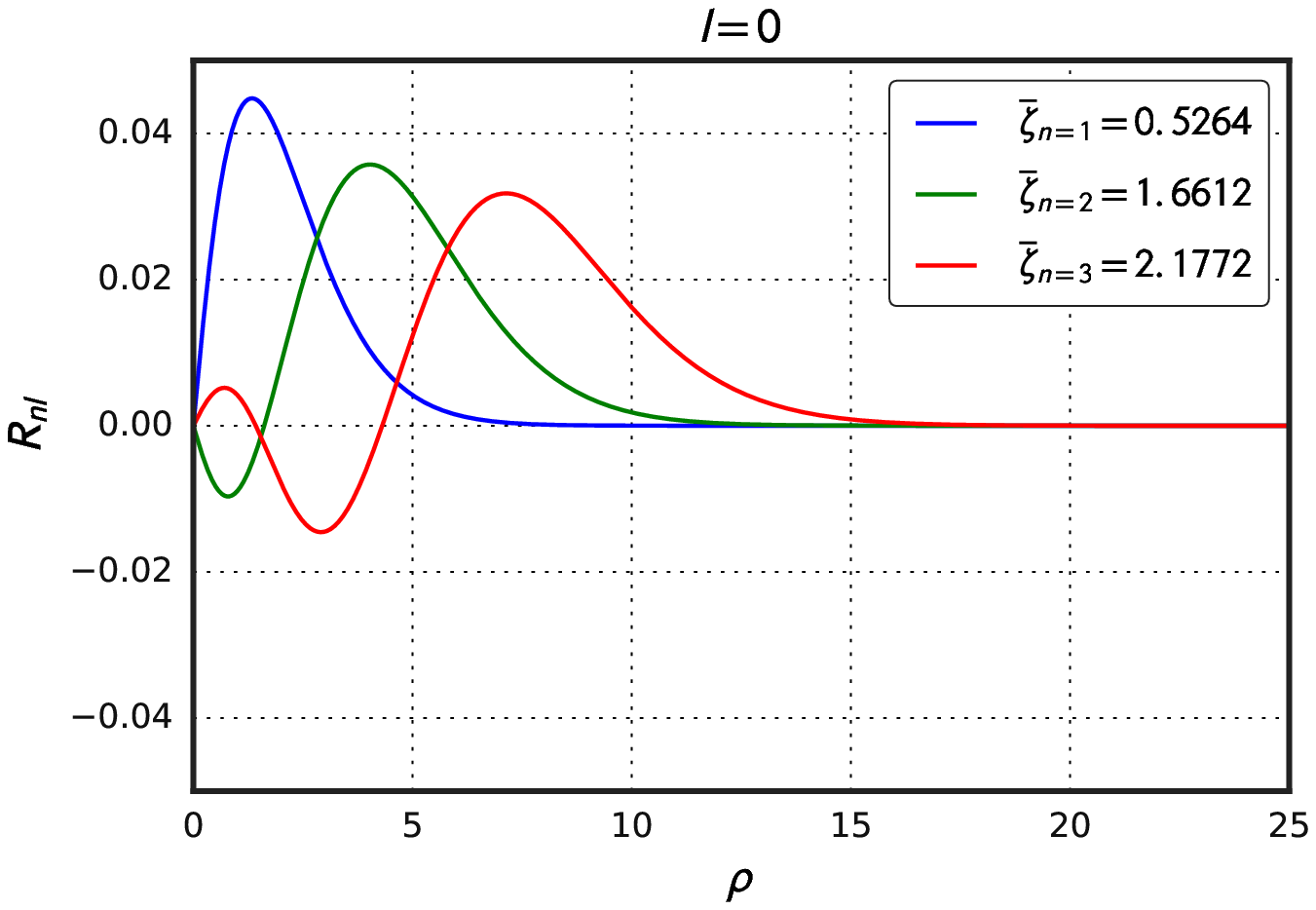} 
	\includegraphics[scale=0.6]{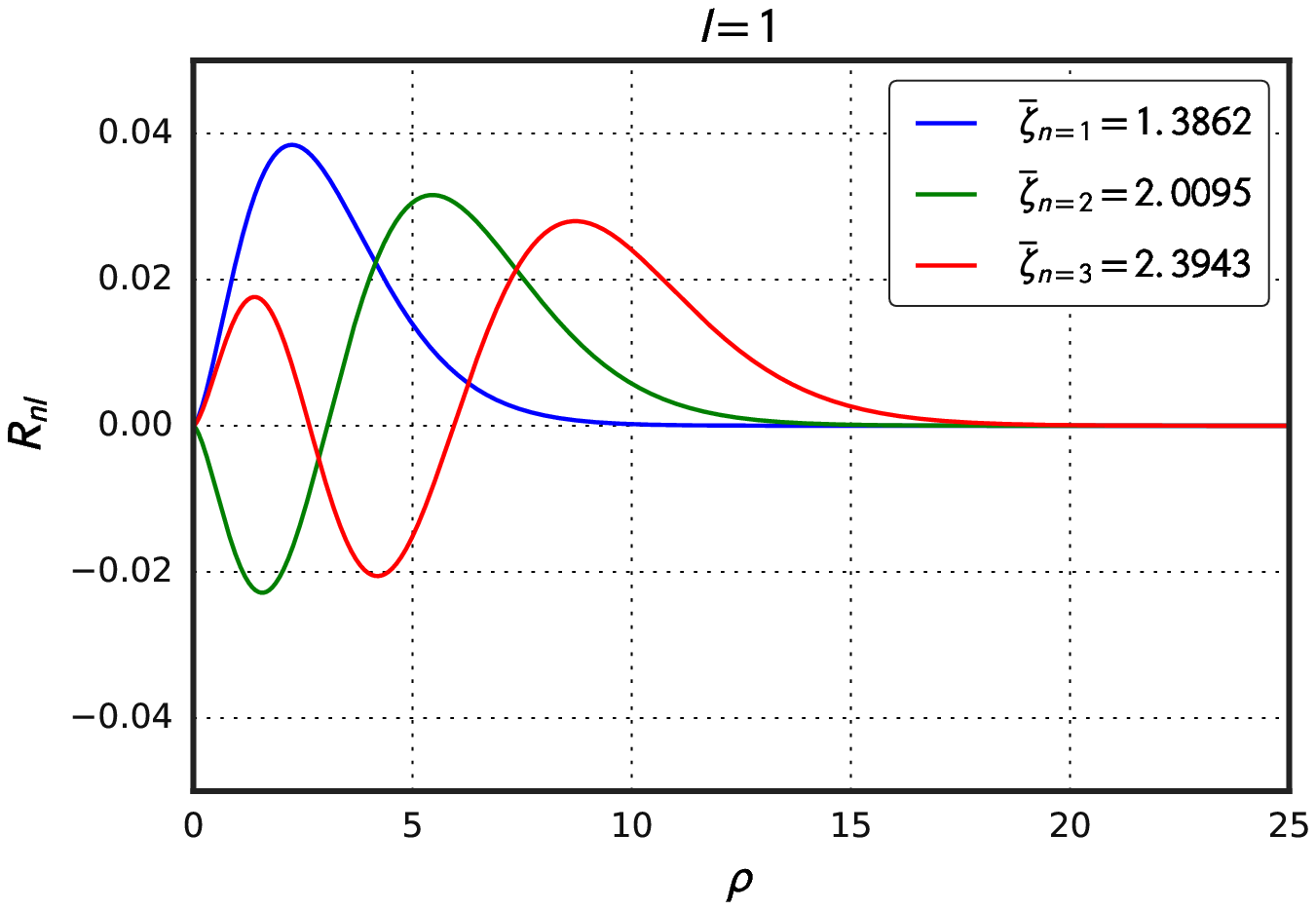} 
	\includegraphics[scale=0.6]{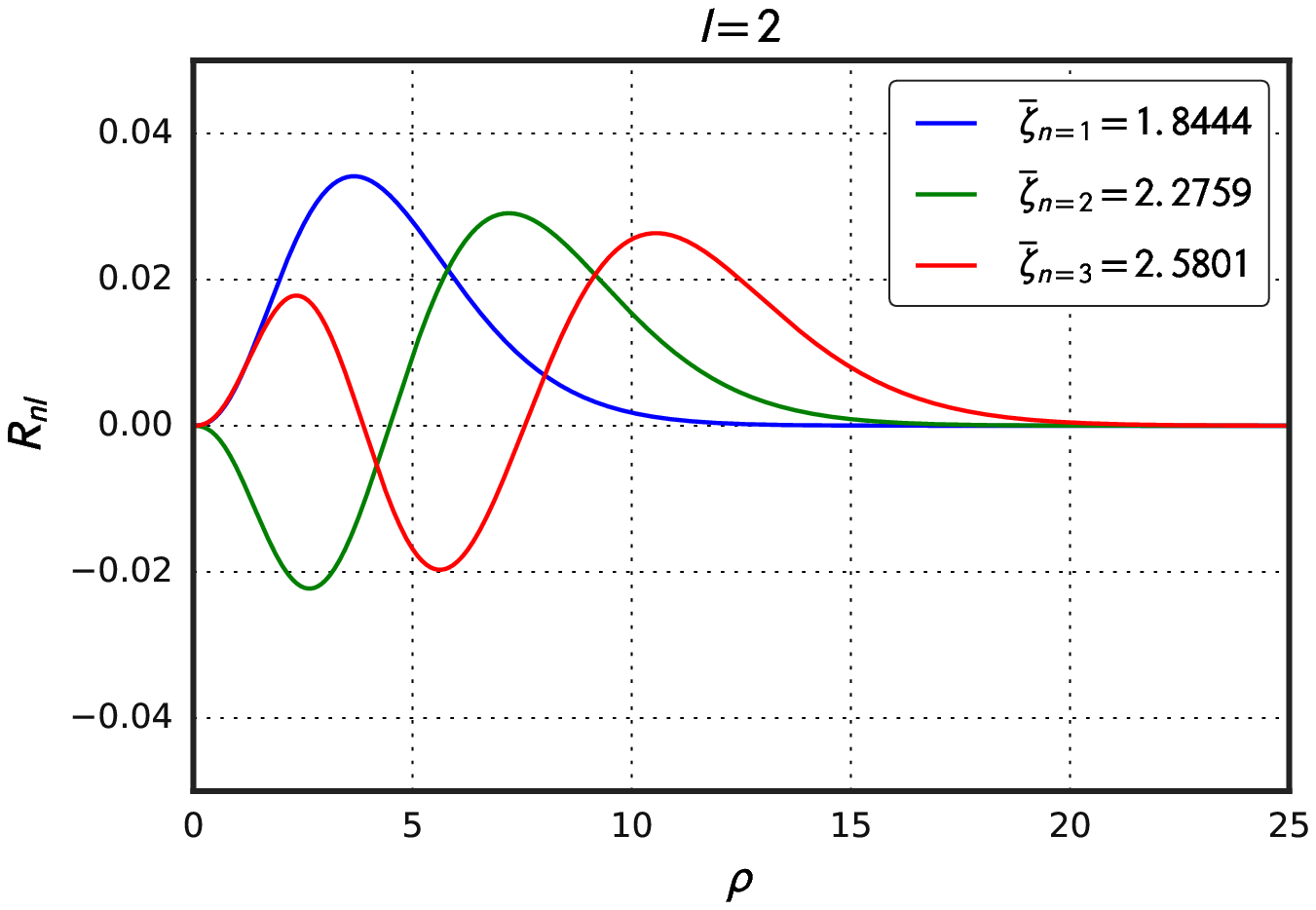}
      \caption{Wave modes for the first lowest eigenvalues $n=1,2,3$ of Eq. (\ref{change}) for each $l=0,1,2$. 
      }
       \label{raw}
\end{figure}

In order to solve equation (\ref{change}), we used a finite difference method \cite{van1996,Franklin2011,burden2011numerical} and computed numerically the radial part of the waves traveling along the wiggly string with their corresponding eigenvalues. The different states are labeled by quantum numbers $n$  (radial quantum number) and $l$. In Fig. \ref{raw}, we plot the lowest three eigenvalues $\bar{\zeta_{nl}}$ of the wave equation (\ref{change}) and the corresponding radial wave amplitudes $R_{nl}(\rho)$ for $l=0,\pm 1, \pm 2$.  

From Eq. (\ref{relation1}) we see that {the} wave modes {that} propagate along the wiggly string axis {are} quantized by $n$ and $l$. It has been suggested that cosmic structures with a non-vanishing Newtonian potential could generally behave as gravitational waveguide for light and massive particles {\cite{Dodonov1989,bimonte1998}} and we examine this proposition in the following. 
The dispersion relation is given by: 
\begin{eqnarray}
\omega^2_{nl}= \frac{1}{n^2_{1}} k^2 \label{drelation34}
\end{eqnarray}
where 
\begin{eqnarray}
n_1=\left[\frac{1-8\varepsilon\bar{\mathcal{\zeta}_{nl}}}{1+8\varepsilon\bar{\mathcal{\zeta}_{nl}}}\right]^{1/2} \label{etaindex}
\end{eqnarray}
is an effective refractive index. 
From Eq. (\ref{drelation34}) we see that modes are propagative along the string provided that the following requirement is fulfilled:
\begin{eqnarray}
0<\bar{\zeta}_{nl} <\frac{1}{4 G(\tilde{\mu}-\tilde{T})}. \label{restri}
\end{eqnarray} 
{This constraint establishes
that the number of wave modes propagating along the wiggly string is large but finite as in an ordinary electromagnetic waveguide.}
As {we would expect}, we find that the allowed {modes}, besides being quantized by $n$ and $l$, {their frequency} also depend on both the energy density and the tension of the string.  

\section{Propagation of Massive Particles}

Since light propagating along a wiggly string is radially confined, as seen in the previous section, it is interesting to investigate what happens to massive particles under the same circumstances. 
In order to study this possibility we write the Klein-Gordon equation ($\hbar=1$) in the wiggly string background geometry:
\begin{eqnarray}
\left[\frac{1}{\sqrt{-g}}\partial_\mu(\sqrt{-g}g^{\mu\nu}\partial_\nu)-m^2\right]\Phi(r,\theta,z,t)=0, \label{Klein}
\end{eqnarray}
where now $\Phi$ is  a complex scalar field describing spinless relativistic particles.
Using the ansatz given in Eq. (\ref{putsolution0}) 
in equation (\ref{Klein}), and  following the same procedures used {above} in the case of massless particles propagation, {we arrive to an identical eigenvalue equation as ({\ref{change}}), but with eigenvalues given}    by 
\begin{eqnarray}
\bar{\mathcal{E}}_{nl}=\frac{1}{8\varepsilon}
\frac{{\omega^2 }-k^2-m^2}{{\omega^2 }+k^2}
\label{relation},
\end{eqnarray}
thus the wavefunctions $R_{nl}(\rho)$ and the eigenvalues $\bar{\mathcal{E}}_{nl}$ are {numerically} identical to the solution of the equation (\ref{change}) (see figure (\ref{raw})). In addition, the discussion on the geometric optics limit of the propagating massless field is still valid for massive particles. However, inclusion of the mass term in the dispersion relation now introduces a cutoff:
\begin{eqnarray}
\omega_{nl}^2=\frac{1}{n^2_2} k^2+\omega_c^2 \label{drelation35}
\end{eqnarray}
where $n_2$ is an effective refractive index defined by
\begin{eqnarray}
 n_2=\left[\frac{1-8\varepsilon\bar{\mathcal{E}}_{nl}} {1+8\varepsilon\bar{\mathcal{E}}_{nl}} \right]^{1/2},
\end{eqnarray}
which {has an identical generic form} than $n_1$, and 
\begin{eqnarray}
\omega_c^2=\frac{m^2}{1-8\varepsilon \mathcal{E}_{nl}} \label{ine}
\end{eqnarray}
is a cutoff frequency. 
\begin{figure}[h] 
	\includegraphics[scale=0.8]{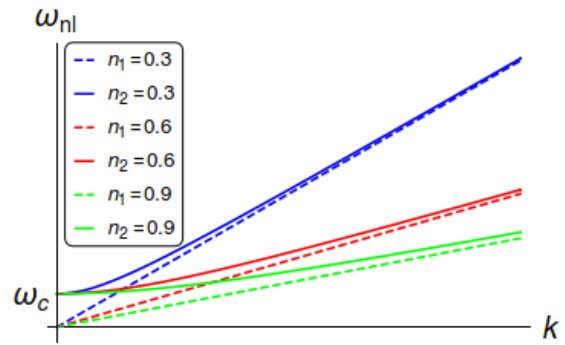} 
      \caption{The angular frequency $\omega_{nl}$ in terms of the wavelength $k$ for different values of  $n_1$ and  $n_2$. While the dashed lines corresponding to massless particles are obviously linear and the solid lines which correspond to the massive case have a quadratic start, both cases are fixed by mode-dependent parameters.}
       \label{refractive} 
\end{figure}
The dispersion relation (\ref{drelation35}) presents a forbidden band as it occurs for electromagnetic waves propagating in an unmagnetized plasma \cite{Jackson1999}. The wave will propagate along the string when its frequency is larger than the cutoff frequency, $\omega_c$, otherwise solutions appear as evanescent waves. At high frequencies, $\omega \gg \omega_c$, we recover the massless dispersion relation (\ref{drelation34}). Moreover,  the constraint
\begin{eqnarray}
0<\bar{\mathcal{E}_{nl}} <\frac{1}{4 G(\tilde{\mu}-\tilde{T})},
\end{eqnarray}
like in the massless case, sets a limit to a finite number  of propagating modes. Besides the dependence on the density of energy and tension of the wiggly string, the allowed propagating modes also depend on the mass of the particle. 

   There is obviously a strong similarity between the propagation of both massless and massive scalar fields along a wiggly cosmic string and the propagation of electromagnetic waves in optical waveguides.  In the next section, we further explore this analogy by proposing a way of designing an optical fiber that mimics the wiggly string in the context described above. 

\section{Analogue Optical Waveguide}

 The analogy between 3D gravity and optics is an old topic that started with the pioneering works of Gordon on Fresnel dragging effect in moving dielectrics \cite{Gordon1923}. Recently, artificial optical materials (metamaterials) have been proposed as a way 
to mimic aspects of curved spacetime in the laboratory. For instance, by manipulating the effective
refractive index  of the medium, Sheng et al. \cite{Sheng2013} were able to reproduce gravitational lensing and trapping of light (see also \cite{Genov2009}).  Incidentally, electronic metamaterials may also be used to simulate peculiar spacetime conditions, like a discontinuous  Lorentzian to Kleinian metric signature change \cite{figueiredo2016modeling} which has also been modeled by optical metamaterials \cite{smolyaninov2010metric}. Liquid crystals, as well, have been used to simulate straight cosmic strings  \cite{satiro2006lensing} and the Schwarzschild spacetime \cite{pereira2011flowing}. On the other hand, it has been proposed that, by spatially varying the doping  concentration, the refractive index profile of optical fibers can be used to control optical transmission in a designer-specified way \cite{Moraes2011}. Following this standpoint, we investigate the proposition of a graded-index optical fiber that reproduces {some of} the properties of the scalar field propagation along a wiggly string. 

{In general, the wave equations for electromagnetic waves propagating along a circular fiber are coupled \cite{Jackson1999}. This implies that there is no separation into purely TE or TM modes but, in the specific case of a fiber with  an azimuthally symmetric refractive index,  if the fields have no dependence on the azimuthal angle,  the equations uncouple into separate scalar wave equations  of the form}
\begin{eqnarray}
\left[\frac{1}{r}\frac{\partial}{\partial r}\left(r\frac{\partial}{\partial r}\right)+\frac{\partial^2}{\partial z^2}+n^2(r)\omega^2\right]\Phi=0, \label{hfiber}
\end{eqnarray}
where $\omega$ is the angular frequency,  $n(r)$ is the optical fiber refractive index and  $\RealP \Phi$ represents any real component of the field. For waves propagating along the optical fiber ($z$-direction), the ansatz {$\Phi(r,z)= e^{-i  k z} R(r)$}, (where $k \in \mathbb{R}$)
substituted  into Eq. (\ref{hfiber}) gives 
\begin{eqnarray}
-\frac{1}{r}\frac{d}{d r}\left(r\frac{dR}{d r}\right)-n^2(r)\omega^2R+k^2R=0. \label{compare1}
\end{eqnarray}
Here, we choose the refractive index to be given by
 \begin{eqnarray}
n(r)=\left(1-  \Omega \ln{\frac{r}{r_0}}\right)^{1/2} ,\label{ourprofile}
 \end{eqnarray}
with the dimensionless parameter $\Omega \ll 1$ in order to be consistent with the wiggly string parameter {$\varepsilon$.} The quantity  $r_0$  is considered to be much smaller than the radius $r_f$ of the fiber and defines an opaque core radius. This way, by considering propagation in the region $r_0 <r<r_f$, the logarithmic singularity at $r=0$ is avoided. Like in the previous sections, we change $r$ to dimensionless units by doing  the change of variables  $\rho=r/\nu$, $\rho_0=r_0/\nu$  and setting {$\nu=\Omega^{-1/2}\omega^{-1}$.} {Then, the  dimensionless equation for the optical fiber can be written as: 
\begin{eqnarray}
-\frac{1}{\rho}\frac{d}{d\rho}\left(\rho\frac{dR}{d\rho}\right)+\left(\ln{\frac{\rho}{\rho_0}}\right)R=\bar{\beta}_{n}R, \label{fibereq}
\end{eqnarray}
where
\begin{eqnarray}
\bar{\beta}_{n}=\frac{1}{ \Omega  }\left({1-\frac{k^2}{\omega^2}}\right).
\label{relation1of}
\end{eqnarray}}

\begin{figure}[h]
	\includegraphics[scale=0.3695]{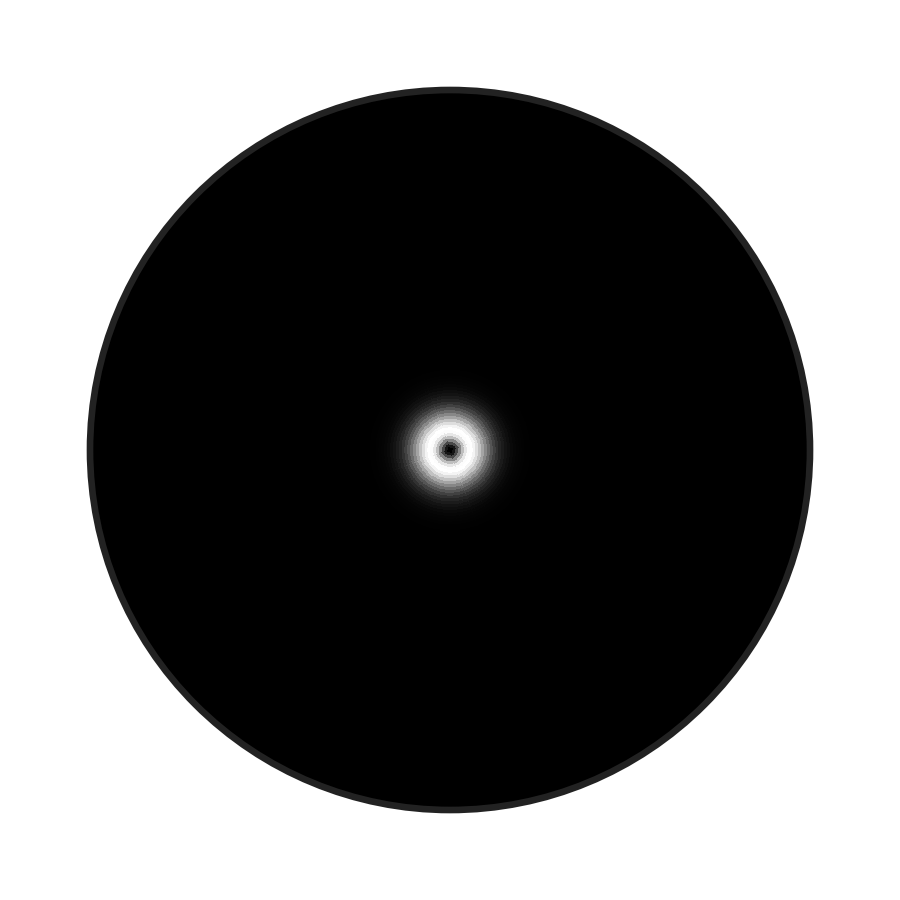} 
    \includegraphics[scale=0.3695]{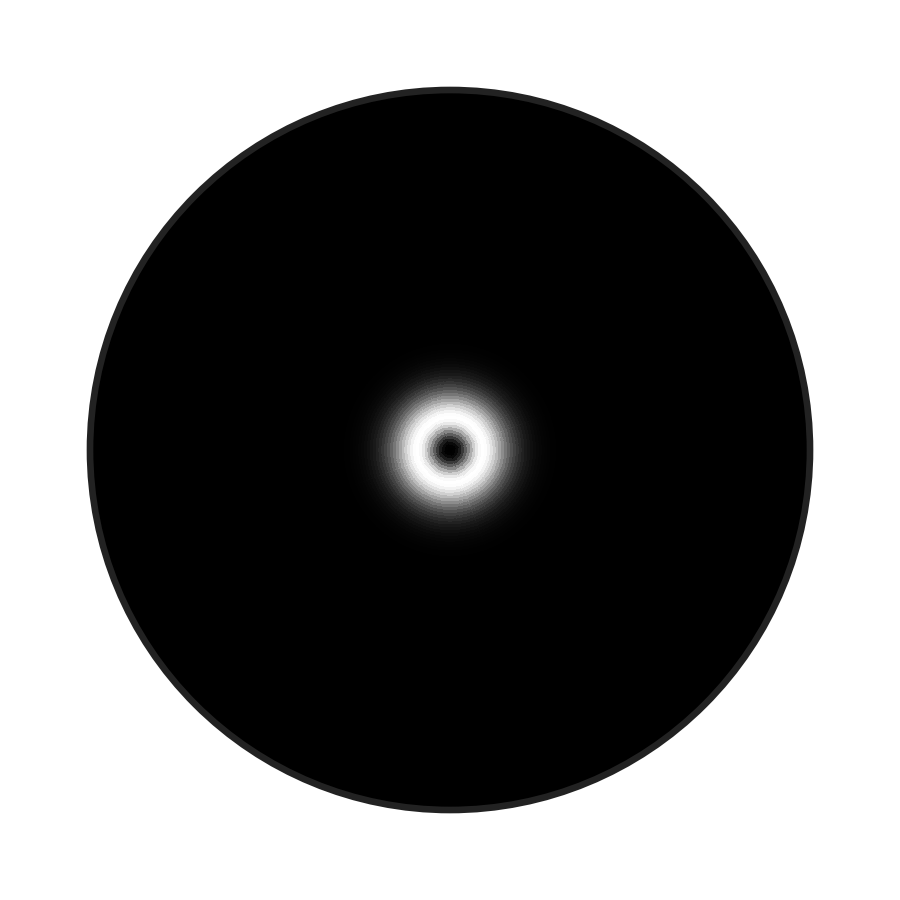} 
    \includegraphics[scale=0.3695]{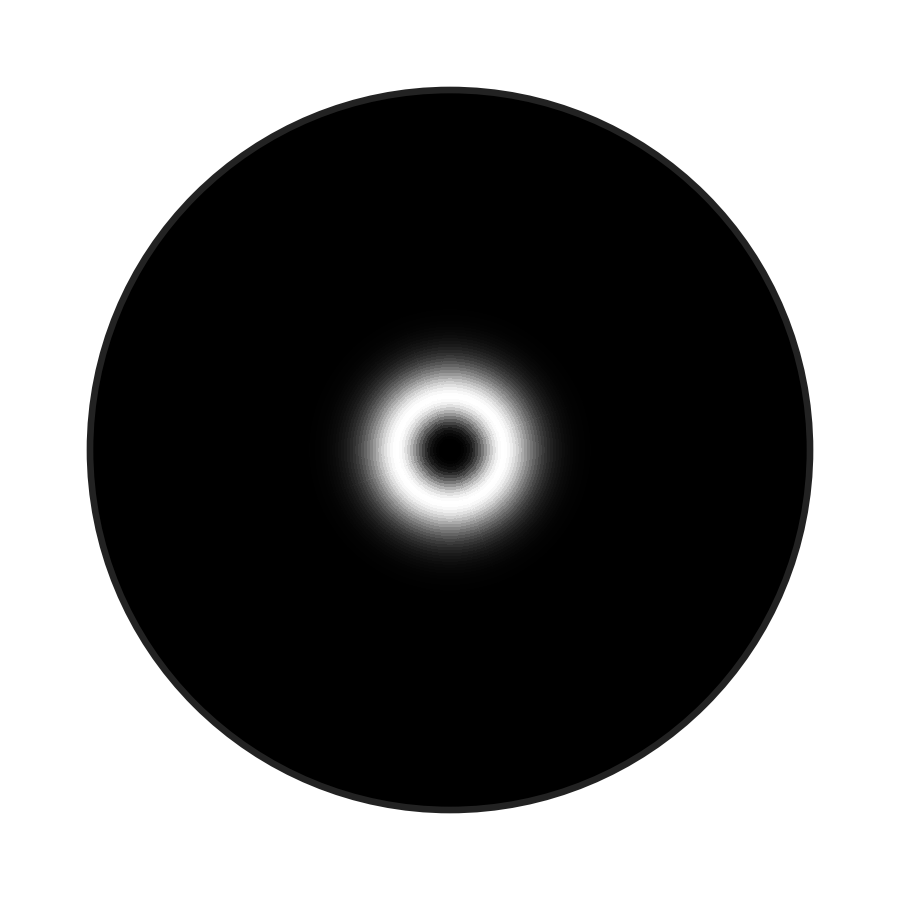} \\
    \includegraphics[scale=0.3695]{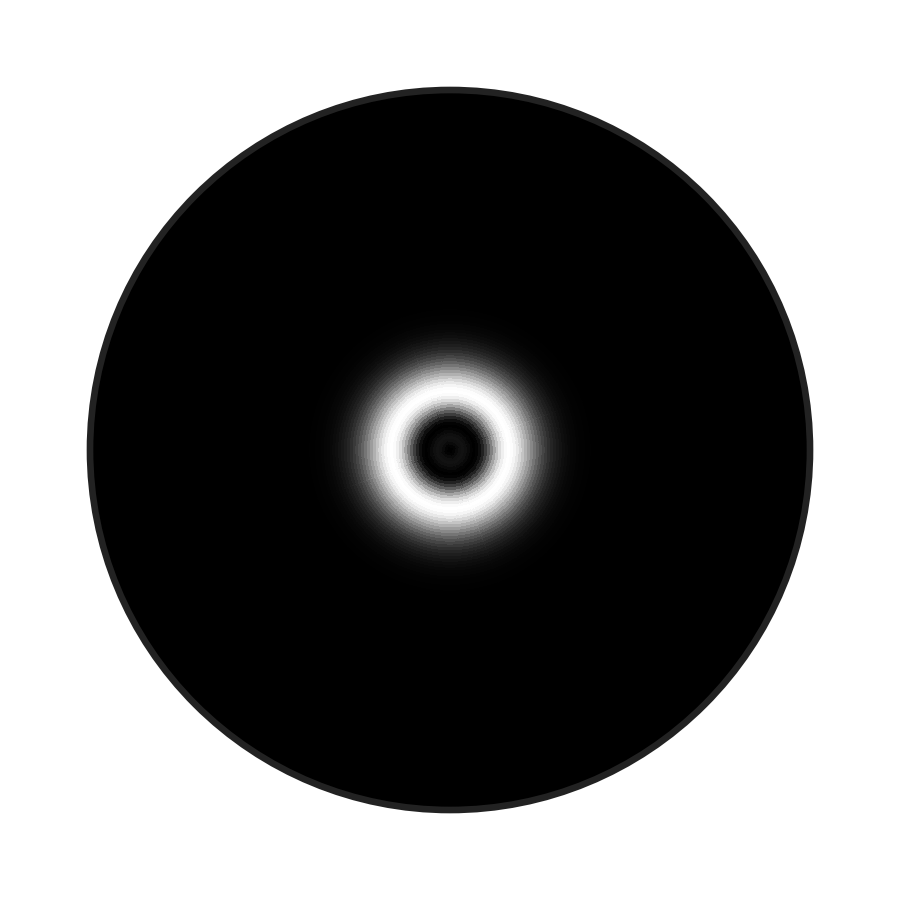} 
    \includegraphics[scale=0.3695]{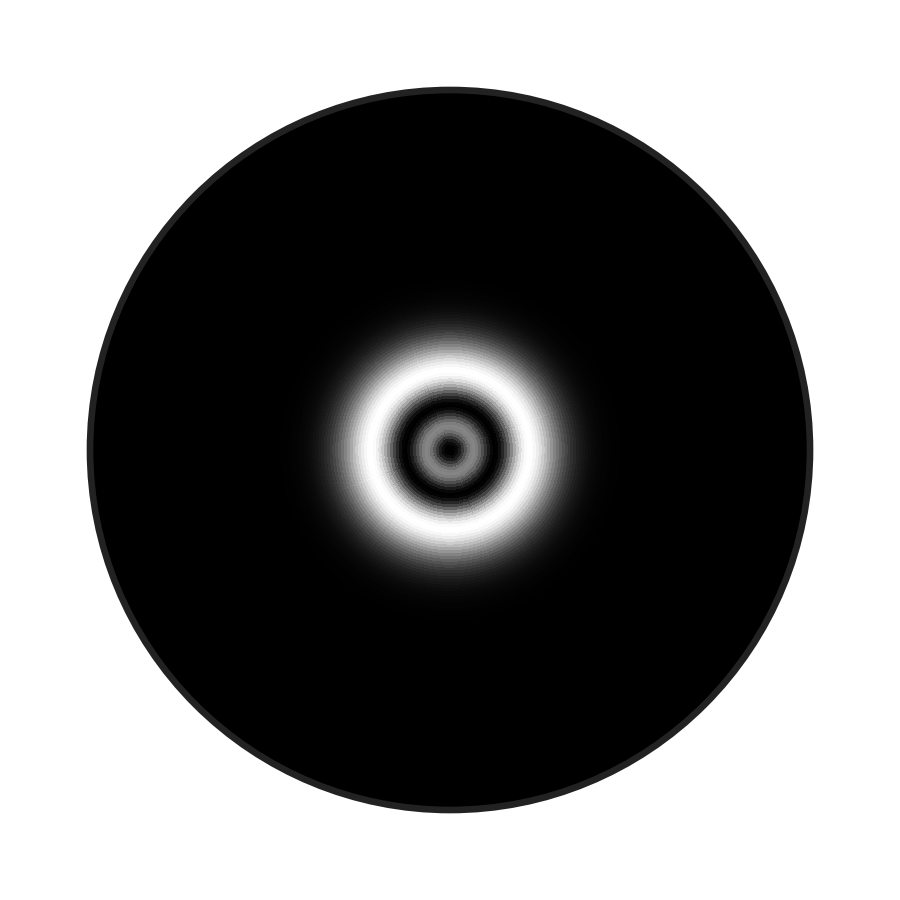} 
    \includegraphics[scale=0.3695]{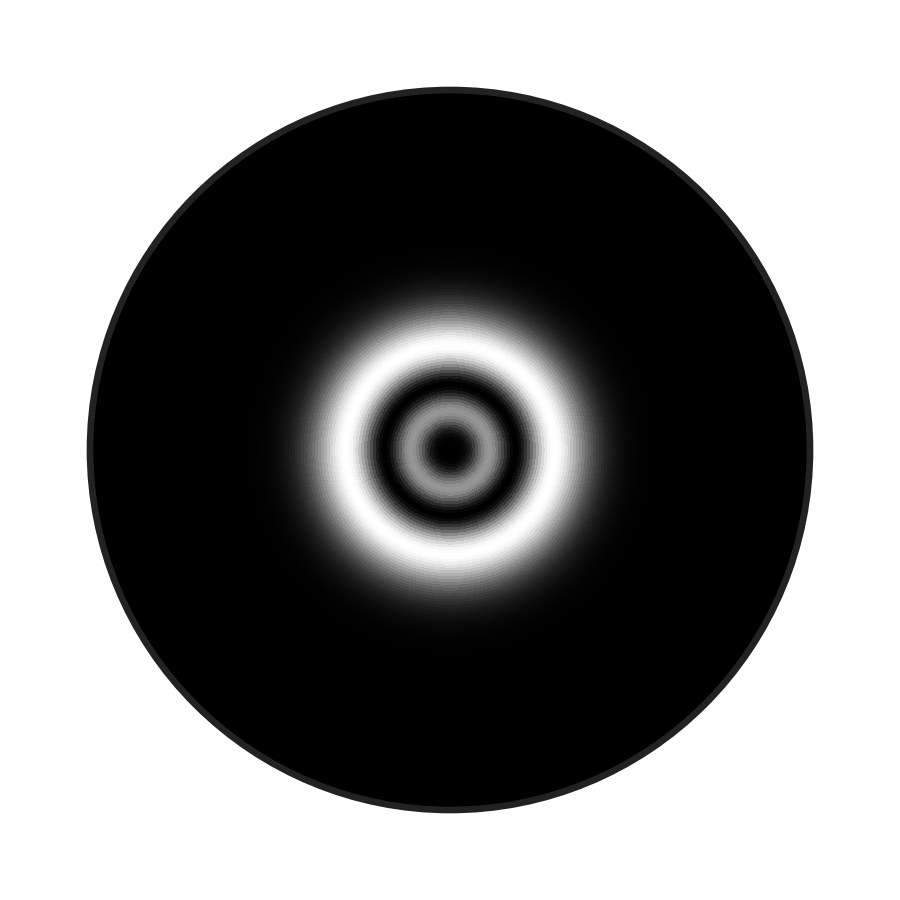}  \\
    \includegraphics[scale=0.3695]{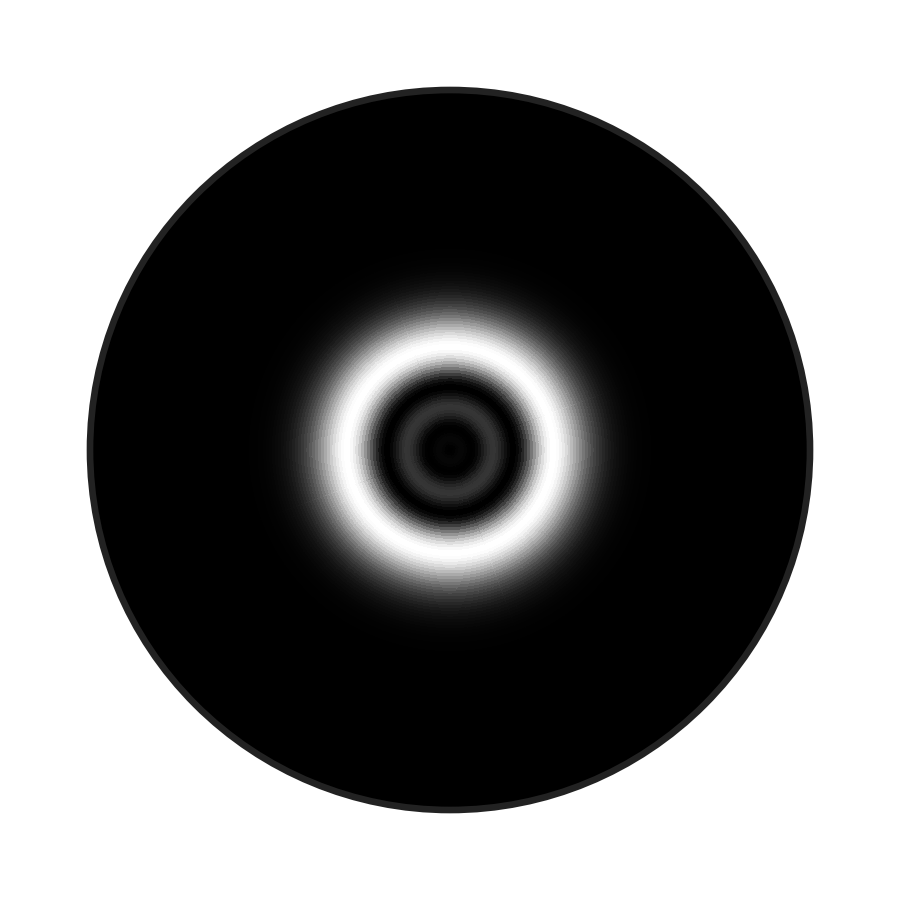} 
    \includegraphics[scale=0.3695]{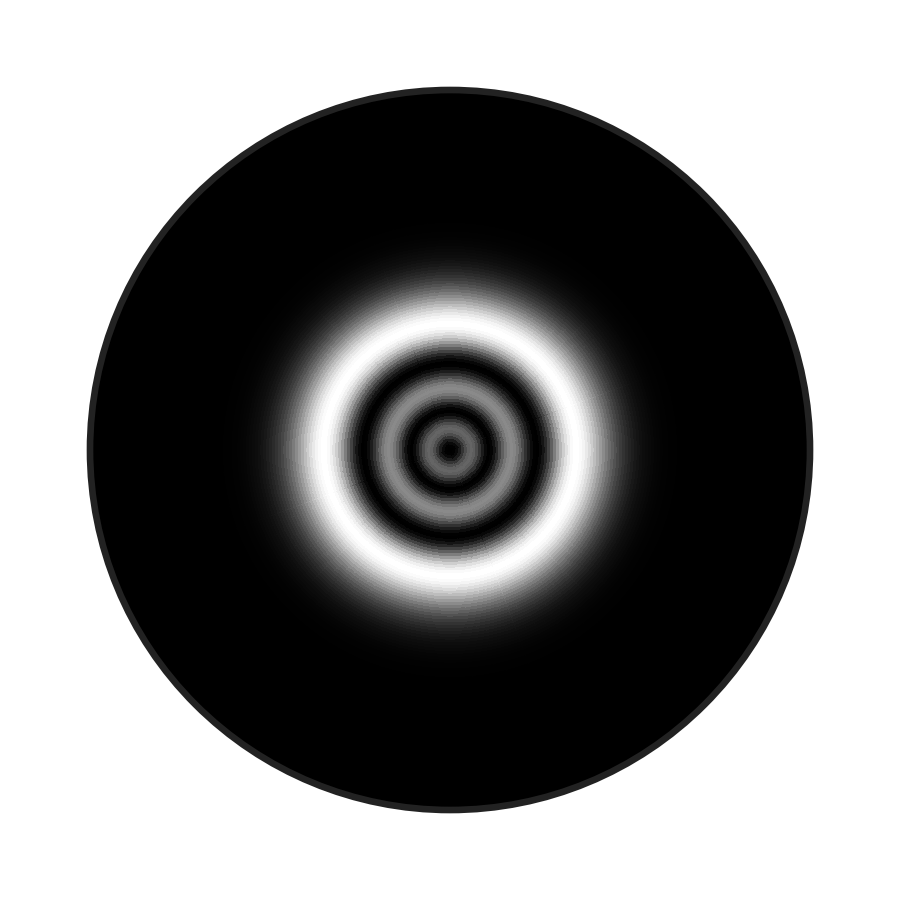} 
    \includegraphics[scale=0.3695]{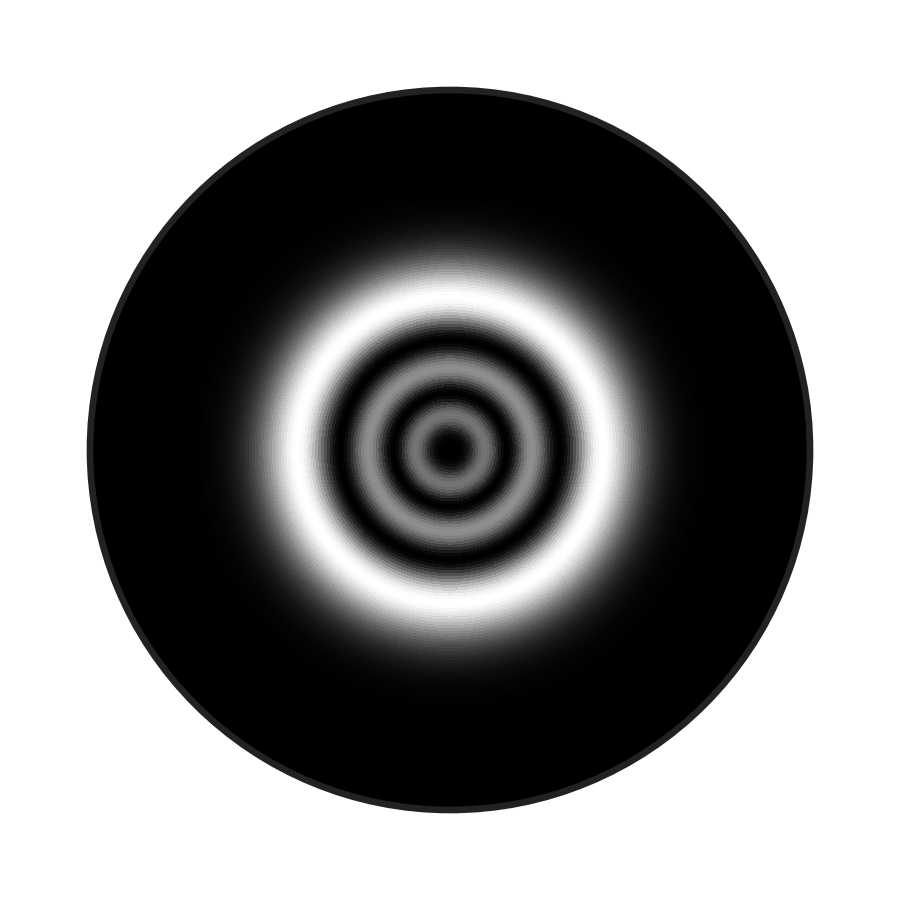}
    \caption{ Intensity profiles {for the first few solutions $R_{nl}^2$ of Eq. (\ref{change})}, with $n=1,2,3$ and $l=0,1,2$. The profiles are disposed as in a matrix where $n$ and $l$ give the line and column number, respectively. {The first column corresponds to the optical fiber modes described by Eq. (\ref{fibereq}). }}
       \label{intensity}
\end{figure}

The radial amplitudes of the wave and their corresponding eigenvalues in the optical fiber with the refractive index given by Eq. (\ref{ourprofile}) obey equations {identical} to {the ones of} massless and massive particles propagating {with $l=0$} in the spacetime of a wiggly string. For a given $z$, the intensity profiles for the propagating waves described by Eq. (\ref{change}) are given by $2\pi \rho R_{nl}^2(\rho)$. In Fig. \ref{intensity}, we plot the intensity distribution for different wave modes, {solutions of Eq. (\ref{change}). The optical fiber modes described by Eq. (\ref{fibereq}) correspond to the cases where $l=0$.}  

Before ending this section, we remark that the coupled equations for the electromagnetic field in the circular optical fiber, in the more general case where the field depends on the azimuthal angle but the refractive index remains azimuthally symmetric, give rise to hybrid HE modes (no longer TE or TM) \cite{Jackson1999}. The case of a step-index fiber was studied in Ref. \cite{snitzer1961cylindrical} which found the field to be of the form $R_l (\rho) e^{il\theta}$, where $R_l$ is a Bessel function. Even though Eq. (\ref{change}) is not a Bessel equation, its symmetries and the shape of the numerical solutions shown in Fig. \ref{raw}, suggest that a solution in terms of an expansion on Bessel functions might be rapidly convergent. The $\theta$-dependence of the scalar fields solutions, for $l \neq 0$, is therefore reminiscent of what happens in the circular optical fiber.  Also, it would be interesting to compare the coupled electromagnetic vector field equations for a wave propagating along a circular optical fiber with refractive index given by Eq. (\ref{ourprofile}) with the electromagnetic field equations in the wiggly string background.  

\section{Conclusion} 

In this paper, we examined the effect of wiggly cosmic strings on propagation of massless and massive fields. We found that waves propagating along the string axis experience the small-scale perturbations which make the propagation qualitatively different from that of waves propagating in the background spacetime with a unperturbed cosmic string. The non-vanishing Newtonian potential acts as an inhomogeneous dielectric medium so that the massless particles are radially confined in a vicinity of the defect axis. Therefore, the wiggly string spacetime behaves as a gravitational waveguide in which wave modes are quantized.  These latter depend on the string energy density and string tension. The number of allowed modes is finite as in a ordinary optical waveguides. On the other hand, the presence of wiggles cause gravitational pullings on massive objects, making the waveguide effect to be also valid for massive fields propagation. In this case, the frequencies of the waves also depend on the mass of the particle.  

Finally, we proposed the design of an optical fiber with a non homogeneous refractive index profile likely to mimic the effect of a perturbed cosmic string. The radial solutions with the corresponding eigenvalues were found by using a numerical method. Although we have considered here the propagation of massive and massless scalar fields along a wiggly string, the extension to vector fields like vector bosons or the electromagnetic field can be of interest. In particular, as a perspective for future work we mention the study of propagating electromagnetic waves along the wiggly string and a possible correspondence with an optical fiber. This is more complex than the problem presented here since the vector field equations are coupled and cannot be reduced to scalar wave equations.  

{Another aspect of of the optical fiber/wiggly string analogy is whether a  propagating  electromagnetic wave along the wiggly string could act as a tractor field on particles in the string vicinity. This has been proposed recently in the realm of negative index optical waveguides \cite{salandrino2011reverse}: instead of being pushed by radiative pressure, a polarizable particle in such environment is attracted by the source of radiation. Even though the requirement of a negative refractive index rules out the wiggly string as such  waveguide, a related  linear defect, the hyperbolic disclination \cite{fumeron2015optics}, seems to be a plausible mediator of this effect. The Kleinian signature of its metric simulates a negative refractive index.}

{Networks of cosmic topological defects have been proposed as models for solid dark matter \cite{bucher1999dark}. This suggests that one might explore the optical implications of a network of wiggly strings. For instance, for a periodic array of strings one might expect some of the properties of a photonic crystal, like the appearance of band gaps in the dispersion relation, which  limit the propagation to the allowed regions of the spectrum. This is presently under investigation and will be the subject of a future publication.}

\acknowledgements
F.M. is grateful to U. Lorraine, FACEPE, CNPq and CAPES for financial support.	F.A. thanks the Coll\`ege Doctoral ``$\mathbb L^4$ collaboration'' (Leipzig, Lorraine, Lviv, Coventry) and the Dionicos programme between U. Lorraine and UNAM for financial support.

\bibliography{references}

\end{document}